\documentclass[pdf,preprint]{iucr}              
\usepackage{graphicx}
\usepackage{url}
     \journalcode{J}              

\begin{document}                  



\title{Lattice strain and tilt mapping in stressed Ge microstructures using X-ray Laue micro-diffraction and rainbow-filtering}


\cauthor[a,b]{Samuel}{Tardif}{samuel.tardif@cea.fr}{}
\author[a,c]{Alban}{Gassenq}
\author[a,c]{Kevin}{Guilloy}
\author[a,c]{Nicolas}{Pauc}
\author[a,d]{Guilherme}{Osvaldo Dias}
\author[a,d]{Jean-Michel}{Hartmann}
\author[a,d]{Julie}{Widiez}
\author[e]{Thomas}{Zabel}
\author[e]{Esteban}{Marin}
\author[e]{Hans}{Sigg}
\author[f]{J{\'e}r{\^o}me}{Faist}
\author[a,d]{Alexei}{Chelnokov}
\author[a,d]{Vincent}{Reboud}
\author[a,c]{Vincent}{Calvo}
\author[a,b,g]{Jean-S{\'e}bastien}{Micha}
\author[a,b]{Odile}{Robach}
\author[a,b]{Fran{\c c}ois}{Rieutord}

\aff[a]{Univ. Grenoble Alpes, 38000, Grenoble, France}
\aff[b]{CEA-INAC-MEM, 17 rue des Martyrs, 38000, Grenoble, France}
\aff[c]{CEA-INAC-PHELIQS, 17 rue des Martyrs, 38000, Grenoble, France}
\aff[d]{CEA-LETI, Minatec Campus, 17 rue des Martyrs, 38054, Grenoble, France}
\aff[e]{Laboratory for Micro- and Nanotechnology, Paul Scherrer Institute, 5232, Villigen, Switzerland}
\aff[f]{Institute for Quantum Electronics, ETH Zurich, 8093, Z{\"u}rich, Switzerland}
\aff[g]{CNRS, 17 rue des Martyrs, 38000, Grenoble, France}









\maketitle                        

\begin{synopsis}
The lattice tilts and full strain tensor are measured in Ge micro-devices under uniaxial or biaxial stress using standard and  rainbow-filtered Laue micro-diffraction. Maps with sub-micron resolution of the strain tensor components are in very good agreement with finite element simulations. 
\end{synopsis}

\begin{abstract}
Micro-Laue diffraction and simultaneous rainbow-filtered micro-diffraction were used to measure accurately the full strain tensor and the lattice orientation distribution at the sub-micron scale in highly strained, suspended Ge micro-devices. 
A numerical approach to obtain the full strain tensor from the deviatoric strain measurement alone is also demonstrated and used for faster full strain mapping.
We performed the measurements in a series of micro-devices under either uniaxial or biaxial stress and found an excellent agreement with numerical simulations. 
This shows the superior potential of Laue micro-diffraction for the investigation of highly strained micro-devices.
\end{abstract}


\section{Introduction}

The development of integrated laser sources compatible with microelectronic technologies is currently one of the main challenges for silicon photonics. 
Since Ge can be CMOS compatible, the interest in strained Ge has significantly increased over the last few years \cite{dutt_roadmap_2012}.
In theory, straining Ge can improve the emission properties by reducing the energy difference between the indirect $L$ valley and the direct $\Gamma$ valley. 
The energy difference is expected to vanish for $4.7\%$ uniaxial or $2.0\%$ biaxial tensile strain \cite{kurdi_band_2010, aldaghri_optimum_2012, sukhdeo_direct_2014}.
Different methods are currently developed to strain Ge layers, either using internal strain redistribution \cite{suess_analysis_2013, gassenq_bi_2015}, or external stress application using stressor layers such as SiN \cite{ghrib_tensile_2013, guilloy_tensile_2015} or mechanical apparatus (e.g. bulge testing) \cite{kurdi_control_2010, sanchez-perez_direct-bandgap_2011, boztug_tensilely_2013}. 
Strain measurements are usually done by micro-Raman spectroscopy \cite{kurdi_control_2010, boztug_tensilely_2013, ghrib_tensile_2013, urena_raman_2013}. 
Raman spectroscopy has the advantage of being relatively well spatially resolved, fast and readily available in laboratories. 
Yet it is mainly limited to the measurement of the Raman spectral shift, which in turn has to be interpreted as a function of the materials, the nature of the strain \cite{cerdeira_stress_1972, wolf_micro_1996} and the light polarization \cite{mermoux_strain_2010}. 
Absolute in-plane strain can also be revealed semi-destructively at micron to millimeter scale using the focused ion beam-digital image correlation technique \cite{lunt_state_2015}.
Further down the spatial scale, high resolution electron backscatter diffraction is another technique that shares some similarities with the technique described in this paper, and that balances limited probing depth with a higher spatial resolution \cite{wilkinson_review_2014}.
Monochromatic synchrotron X-ray micro-diffraction measurements were also performed in Ge devices \cite{ capellini_strain_2013, etzelstorfer_scanning_2014, ike_characterization_2015, chahine_strain_2015, keplinger_strain_2016}.
Direct measurements of the atomic planes spacing with sub-micron resolution were obtained. 
However several orientations had to be measured successively in order to resolve the full strain tensor and the different probing volumes of the different orientations further complicated the analysis. 

Laue micro-diffraction is well suited for strain and orientation mapping in miscellaneous materials, is compatible with \textit{in situ} mechanical tests and can be correlated with other techniques to reveal details about plastic deformation \cite{chung_automated_1999, tamura_strain_1999, ice_3d_2000, macdowell_submicron_2001, larson_three_2002, tamura_high_2002, tamura_submicron_2002, tamura_scanning_2003, rogan_direct_2003,ice_polychromatic_2005, maass_defect_2006, kunz_dedicated_2009, ice_tutorial_2009, huang_cyclic_2009, kirchlechner_dislocation_2011, kirchlechner_in_2011, robach_full_2011, villanova_multiscale_2012, chen_unambigous_2012, richard_strains_2012, korsunsky_analysis_2012, hofmann_x-ray_2013, ibrahim_laue_2015, dejoie_complementary_2015, kirchlechner_reversibility_2015}.
Recently, local Laue micro-diffraction measurements of the strain have also been performed in nano-objects, such as Au nanowires \cite{leclere_in_2015} or Ge nanowires \cite{guilloy_tensile_2015}. 
However, the nature of the white beam limits the measurement to the deviatoric strain, which needs to be interpreted to access the lattice parameter value. 

Since the initial works on the Laue micro-diffraction technique, it was recognized that the additional ability to measure the energy of at least one Bragg reflection would enable the determination of the full strain tensor (see \textit{e.g.} \cite{chung_automated_1999, tamura_strain_1999}). 
Robach \textit{et al.} \cite{robach_tunable_2013} have recently demonstrated that this energy measurement could be performed in a white beam configuration, \textit{i.e.} in the absence of a monochromator, using the so-called ``rainbow-filter'' technique.
This technique is akin to a reverse monochromatic filtering where specific energies are removed from the incident spectrum using a diamond plate in diffraction conditions upstream of the sample. 
Such method combines both advantages from poly- and monochromatic micro-diffraction, \textit{i.e.} a complete determination of both sample orientation and strain. 

Here we report on rainbow-filtered Laue micro-diffraction measurements of the full strain tensor in Ge microstructures, in which the strain is redistributed and concentrated in a small region. 
We first measured the deviatoric strain and the lattice tilts distribution in uniaxially and biaxially stressed micro-devices using Laue micro-diffraction, thus unambiguously separating the tilt and strain contributions. 
We then obtained the full strain tensor in the center region of the microstructures using the rainbow-filter technique and confirmed the absence of normal stress on the free surfaces. 
The latter observation was used to numerically retrieve the full strain tensor from (standard) Laue micro-diffraction measurements alone. 
We could thus further extend the full strain tensor determination to the complete mapping of the micro-devices. 
Our work demonstrates the high potential of standard and rainbow-filtered Laue micro-diffraction for the investigation of inhomogenously strained microstructures.

\section{Materials and methods}
\subsection{Sample preparation}
\begin{figure}\label{Fig1}
\caption{
(a) Process flow used for the fabrication of strain-redistributed microstructures from $\left(001\right)$ optical GeOI wafers.
After the under-etching step, membranes are suspended and the large pads relax, concentrating the strain in the center region of the bridges ; general designs and critical parameters used to tune tensile strain in the Ge microstructures (b) for uniaxial stress concentration (c) for biaxial stress concentration.}
\includegraphics[width=1.\textwidth]{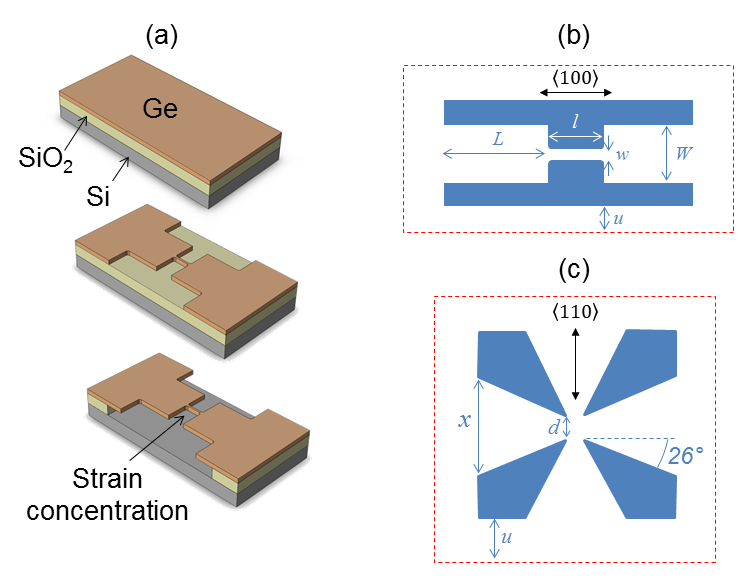}
\end{figure}

Ge suspended micro-bridges and micro-crosses were processed in optical GeOI substrates with a small built-in biaxial tensile strain ($0.16\%$ at room temperature) \cite{reboud_structural_2015, reboud_ultra-high_2016}. 
This initial strain was concentrated in smaller regions using the approach initially proposed for Si micro-bridges \cite{minamisawa_top_2012}: the $0.35$~$\mu m$-thick Ge layer was patterned in the shape of micro-bridges or micro-crosses with large stretching pads on each side using e-beam lithography followed by dry etching in an inductively coupled plasma reactor. 
The membrane is released by etching the underlying oxide using vapor HF and ethanol. This results in strain relaxation in the large pads and a larger biaxial or uniaxial tensile strain is imposed to the small center region (Figure \ref{Fig1}(a)). 
The strain state in the center region can be tuned by the different design parameters, e.g. number of pads, size of the pads and of the central region, as well as the relative orientations of the micro-structure and of the crystal axis.
A detailed studies of their influence has been presented elsewhere \cite{suess_analysis_2013, gassenq_bi_2015}. 
Figure \ref{Fig1} presents the studied designs for $\left\langle100\right\rangle$ uniaxial (Figure \ref{Fig1}(b)) or $\left(001\right)$ biaxial (Figure \ref{Fig1}(c)) stress concentration. 

\subsection{Finite Elements Method simulations}
The strain redistribution was simulated using the 2 dimensional Finite Elements Method (FEM) with the COMSOL Multiphysics software. 
The germanium second order stiffness parameters were used to account for the crystal anisotropy and the elastic coefficients were taken from Ref. \cite{levinshtein_handbook_1996}. 
An in-plane stress corresponding to the $0.16 \%$ initial strain in the GeOI and a zero normal stress were applied to the layer. 
We defined a contour around the micro-bridge or micro-cross corresponding to the under-etching front (dashed lines in Figure \ref{Fig1}(b,c), $u=80~\mu m$). 
Outside this contour, the displacement of the Ge layer was set to zero, since it is clamped by the underlying and much larger Si substrate, while inside the contour the suspended Ge membrane was free to elastically relax. 
For comparison with strain mapping measurements hereafter, the FEM simulations were convoluted with a Gaussian function ($0.8~\mu m$ full width at half-maximum) corresponding to the size of the measurement probe.

\subsection{Rainbow filtered Laue micro-diffraction}
The Laue micro-diffraction experiments were performed at beamline BM32 of the European Synchrotron Radiation Facility \cite{ulrich_new_2011}. 
The incoming white beam (5-25~keV) was focused using non-dispersive Kirkpatrick-Baez mirrors to a sub-micron spot, typically $0.5~\mathrm{\mu m}$ (horizontal) x $0.5~\mathrm{\mu m}$ (vertical).
Due to the $40^\circ$ tilt of the sample surface with respect to the horizontal plane, the vertical footprint of the beam on the sample was about $0.8~\mathrm{\mu m}$, however. 
The sample was positioned in the beam using a three-dimensional stage with 100~nm resolution. The relative position of the sample and the X-ray spot was monitored using a multi-channel analyzer fluorescence detector set to the Ge K$_\alpha$ fluorescence line. 
The Laue diffraction patterns from the sample were collected on a $2048 \times 2048$ pixel MarCCD charge-coupled device (CCD) sensor with an effective pixel size of $80~\mathrm{\mu m}$, positioned 70 mm above the sample. 
Accurate position and angles of CCD pixels with respect to sample and incoming beam directions were determined by the standard calibration procedure on a stress-free Ge thick wafer \cite{chung_automated_1999}.
The orientation matrix and the deviatoric strain tensor were calculated from the peak positions using the LaueTools software \cite{lauetools}. 
We further used the so-called “rainbow-filter” technique to measure the energy of the Bragg reflections and therefore the lattice parameters \cite{robach_tunable_2013}. 
A $300~\mathrm{\mu m}$-thick diamond plate was inserted in the X-ray beam such that energies corresponding to diffractions by the diamond plate were attenuated downstream of the plate.
Rotating the diamond plate results in changing the diffraction conditions so that a set of attenuated transmitted energies could be scanned across the beam spectrum.
After calibration of the energy filter with a known material (here bulk $\left\langle111\right\rangle$ germanium), the relationship between the angular position of the diamond plate and the attenuated energies could be established. 
The energy of any Bragg reflection could then be accessed by scanning the diamond plate while measuring the intensity of the Bragg reflection and correlating the observation of a reduction in intensity with the angular position of the diamond plate. 

\section{Results and discussion}

\subsection{Laue diffraction pattern measurements} 

\begin{figure}\label{Fig2}
\caption{
(a) SEM micrographs of a $\left\langle100\right\rangle$ micro-bridge showing the design parameters $W$ ($51~\mu m$), $w$ ($1~\mu m$), $L$ and $l$ ($11~\mu m$). $L$ was tuned between $25$ and $225~\mu m$ to change the strain amplification. The direct space lattice vectors are indicated by a, b and c.
(b) Raw Laue pattern, as measured in the center region of the micro-bridge and (c) details on two different Bragg peaks, showing the diffraction from both the Ge micro-bridge and the underlying Si substrate as a function of the tensile stress.}
\includegraphics[width=1.\textwidth]{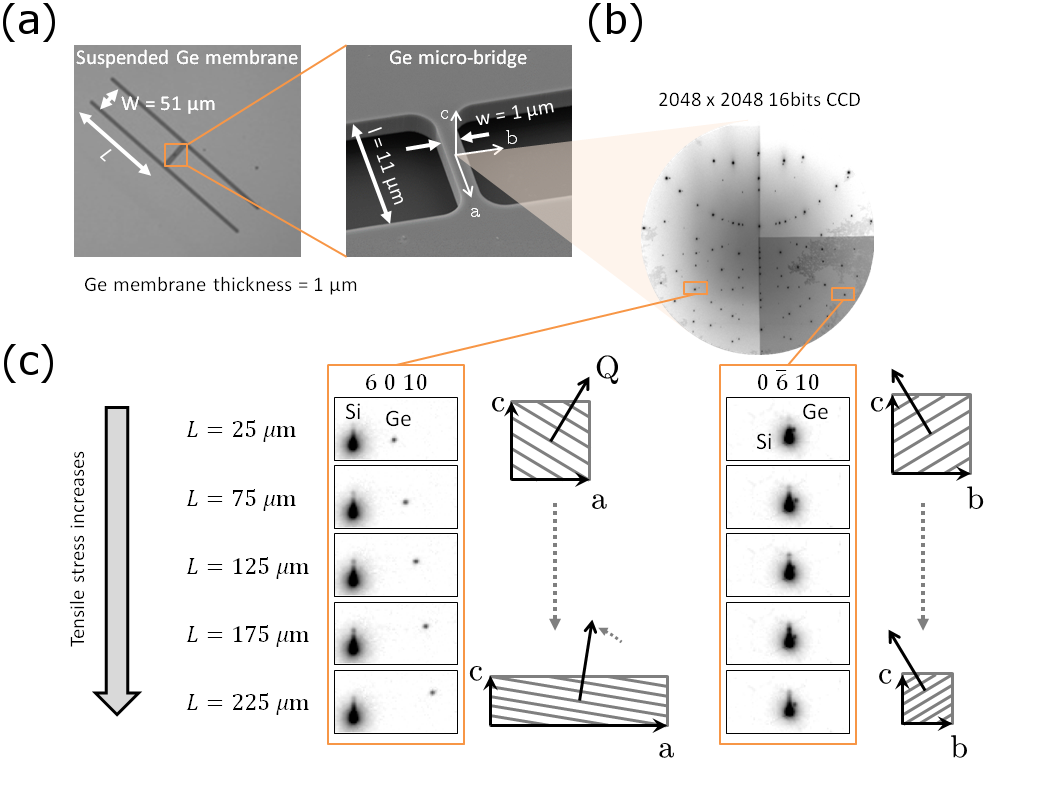}
\end{figure}

Typical Laue patterns measured in the center region of a series of micro-bridges are shown in Figure \ref{Fig2}.
All the bridges were patterned on the same chip in a $0.35~\mathrm{\mu m}$ Ge layer thickness with $11~\mathrm{\mu m}$ bridge length along the $a$ axis and $1~\mathrm{\mu m}$ width ($l=11~\mathrm{\mu m}$, $W=51~\mathrm{\mu m}$ and $w=1~\mathrm{\mu m}$ in Figure \ref{Fig1}(b)).
Different lengths $L$ of the stretching pad were measured, corresponding to different strain states \cite{minamisawa_top_2012}. 

The positions of the $h0l$ peaks (e.g. $6~0~10$ in Figure \ref{Fig2}) vary monotonically with the strain: the strains along $a$ and $c$ have opposite signs due to the Poisson effect, thus the scattering vector rotates around the $b$ axis as the strain increases. 
As a result, the scattering plane rotates as well and the position of the reflected beam moves on the detector. 
On the opposite, the positions of the $0kl$ peaks (e.g. $0~\bar{6}~10$ in Figure \ref{Fig2}) do not change with the strain state since the symmetry is preserved in the $(b,c)$ plane. 
Due to the small Ge thickness, both Ge and Si Laue patterns can be observed. 
Note that the circular shape of the Ge diffraction peaks is indicative of (\textit{i}) a good homogeneity of strain and orientation, and (\textit{ii}) a lack of plastic relaxation at the beam location for the samples probed.
The small offset between the Si and Ge peaks comes from the small misalignment of the two wafers at the moment of their bonding during GeOI fabrication \cite{reboud_structural_2015}, typically lower than $0.25^\circ$.

\begin{figure}\label{Fig3}
\caption{
Raw data from the MarCCD detector used in the Laue diffraction setup in beamline BM32 of ESRF and associated peak positions (Ge: red circles, Si: black rectangles). 
Measurements were performed (a) in the center region of a Ge bridge strained at $3.7\%$ and (b) in the Si substrate close to the bridge; (c) image subtraction between (a) and (b); (d) (e) and (f): corresponding zooms on the diffracted peaks $337$ and $226$.}
\includegraphics[width=1.\textwidth]{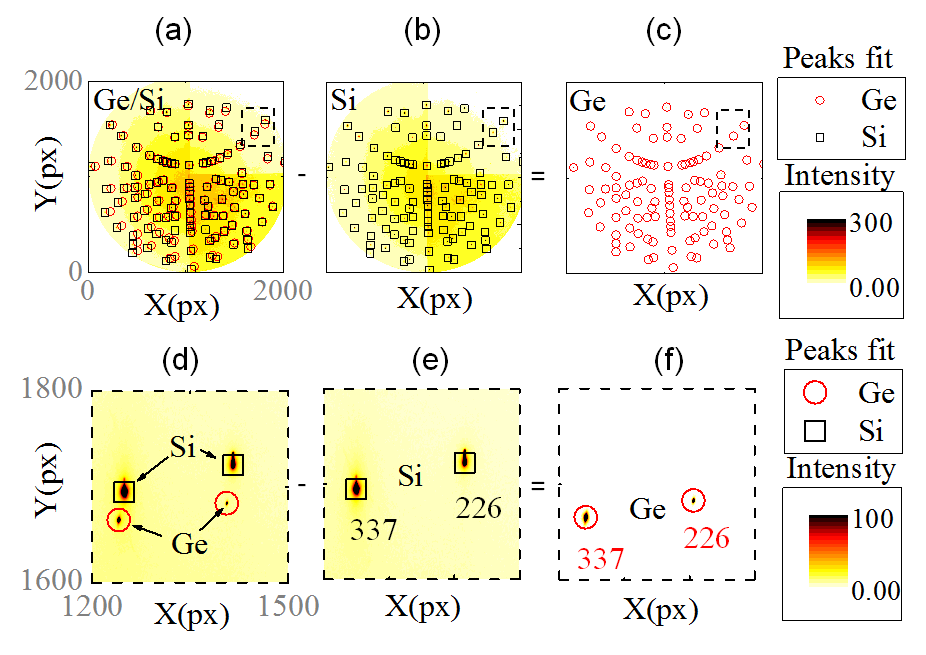}
\end{figure}

Figure \ref{Fig3} presents a typical diffraction pattern measured on the MarCCD detector. 
The associated fits of the maximum intensity of the diffracted peak are indicated by circles for Ge and rectangles for Si. 
In order to extract only the Ge peaks from the diffraction pattern measured in the center region of the bridge (Figure \ref{Fig3}(a)), the background was also measured in the Si substrate close to the bridge (Figure \ref{Fig3}(b)). 
The Si peaks were then removed by subtracting the two images (Figure \ref{Fig3}(c)). 
Figure \ref{Fig3}(d), (e) and (f) are zoom-in on the $337$ and $226$ diffraction peaks, where it clearly appears that the Si peaks are removed by the background processing.

\subsection{Lattice distortions and tilts mapping}
\begin{figure}\label{Fig4}
\caption{
Crystallographic angles $\alpha$, $\beta$ and $\gamma$ and lattice orientation angles $R_x$, $R_y$ and $R_z$ measured in a micro-bridge. 
Inset: orientation of the basis and direct space lattice vectors, and sketch of the strained micro-bridge.}
\includegraphics[width=1.\textwidth]{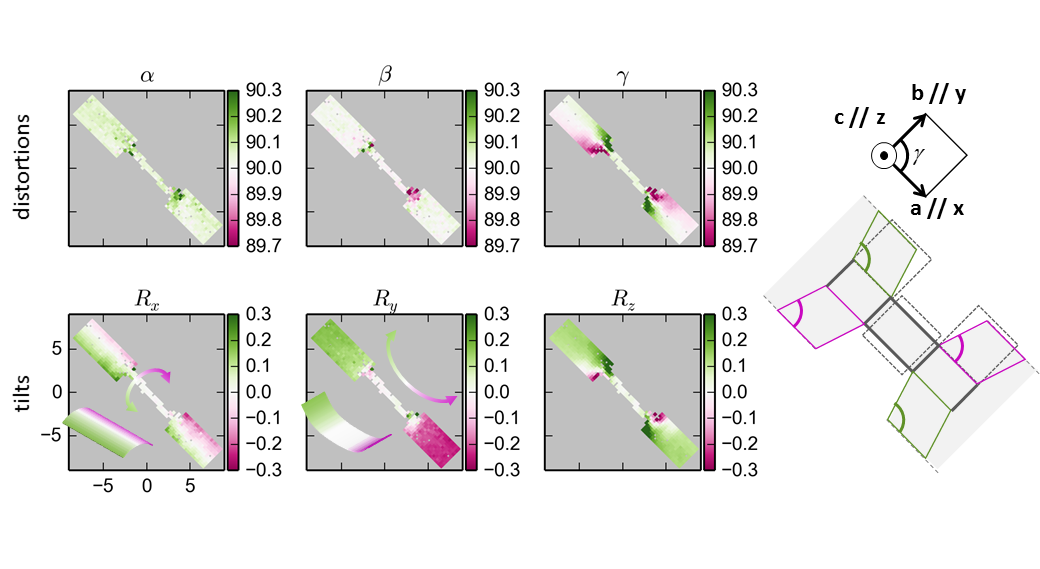}
\end{figure}

\begin{figure}\label{Fig5}
\caption{
Crystallographic angles $\alpha$, $\beta$ and $\gamma$ and lattice orientation angles $R_x$, $R_y$ and $R_z$ measured in a micro-cross. 
Inset: orientation of the basis and direct space lattice vectors, and sketch of the strained micro-cross.}
\includegraphics[width=1.\textwidth]{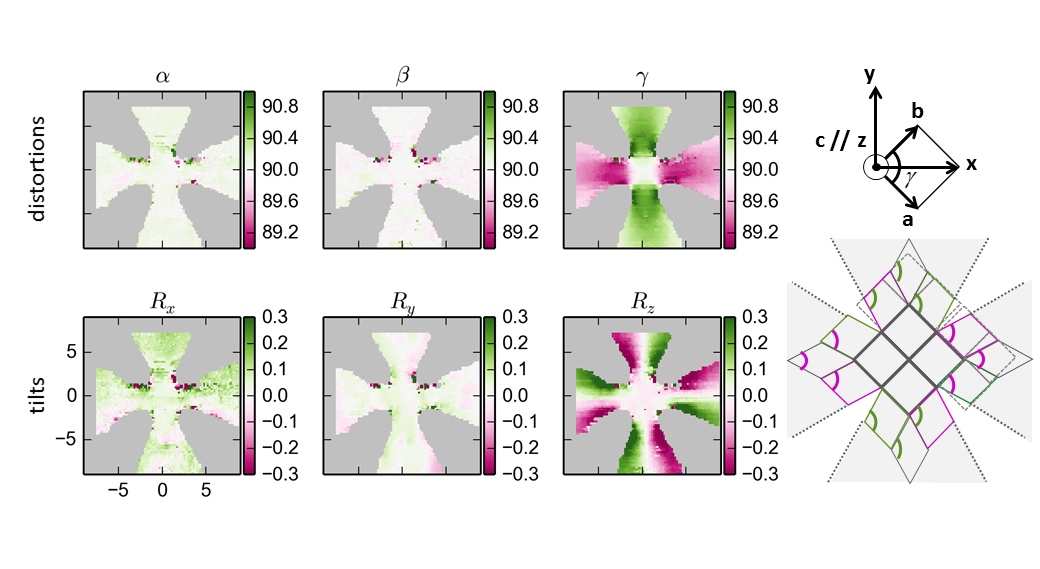}
\end{figure}

On each Laue pattern, up to 80 of the most intense and well-defined peaks were fitted using LaueTools with 8 independent parameters: 5 for the deviatoric strain tensor and 3 for the lattice orientation matrix. 
Figure \ref{Fig4} and \ref{Fig5} present maps of the crystallographic angles and lattice tilts for the uniaxial design ($W=51~\mathrm{\mu m}$, $w=0.5~\mathrm{\mu m}$, $L=75~\mathrm{\mu m}$, $l=12~\mathrm{\mu m}$ and $1~\mathrm{\mu m}$ fillet radius in Fig. \ref{Fig1}(b)) and the biaxial design ($x=140~\mathrm{\mu m}$, $d=7~\mathrm{\mu m}$ and $1~\mathrm{\mu m}$ fillet radius in Fig. \ref{Fig1}(c)), respectively. 
The three crystallographic angles $\alpha$, $\beta$ and $\gamma$ describe the distortions of the cubic lattice while the three rotation angles $R_x$, $R_y$ and $R_z$ are calculated from the lattice orientation matrix and describe the rotations of the lattice around the $x$, $y$ and $z$ axis indicated in the figures.  
In the case of the micro-bridge (Fig. \ref{Fig4}), the in-plane lattice angle $\gamma$ increases over $90^\circ$ or decreases under $90^\circ$ by a few tenths of a degree on each side of the constriction, near the fillets. 
At the same locations, an in-plane rotation of the lattice of similar amplitude is observed in the $R_z$ map. 
The combination of these two effects is illustrated by the simple sketch in Fig. \ref{Fig4}: the lattice is strongly deformed by shear strain along $x$ on each side of the constriction at the position of the fillets, where the uniaxial symmetry is broken (the width of the pad is much larger than that of the bridge). 
Such lattice distortion may cause fractures in the fillet region and can be overcome using grade optimization \cite{nam_strain_2013}. 
The out-of-plane rotations $R_x$ and $R_y$ also display an interesting behavior: while the lattice tilts are negligible in the center region, a rotation around the bridge axis ($R_x$) as well as a rotation perpendicular to the bridge axis ($R_y$) are clearly observed in the pads. 
The $R_x$ and $R_y$ curvatures are sketched in the insets of Fig. \ref{Fig4} and their opposite signs indicate an anticlastic deformation of the micro-bridge, consistently with previous results from Etzelstorfer \textit{et al.} \cite{etzelstorfer_scanning_2014}. 
We note that in our case, the symmetry breaking between the top and bottom surfaces may be due to remaining unetched layers or patches of Si (used as an oxidation protection layer) on the bottom surface of the Ge stretching arms.

Similarly, the stretching pads of the micro-cross display particular lattice tilts and distortions (Fig. \ref{Fig5}). 
The lattice angle $\gamma$ is remarkably larger than $90^\circ$ in the vertical ($y$-axis) pads and smaller than $90^\circ$ in the horizontal ($x$-axis) pads while the in-plane rotation angle $R_z$ changes sign between the edges of each pad. 
Such behavior is different from the micro-bridge case where $\gamma$ and $R_z$ were correlated. 
This effect can be illustrated by the simple sketch in Fig. \ref{Fig5}: since the $x$-axis and $y$-axis pads are here aligned along $\left[110\right]$ and $\left[\bar{1}10\right]$ respectively, the $\gamma$ angle will be reduced (increased) when the lattice is stretched along $x$ ($y$). 
Near the pad edges, a small lattice rotation occurs to accommodate the changes in symmetry, mirrored along the center axis of the pads and thus changing sign across the pad width. 
Accordingly, a vanishing rotation is observed along the center axis of each pad. 
No significant out-of-plane lattice tilts are measurable.
Finally, it can be seen that the orthorhombic symmetry of the lattice is well preserved in the center region of both the micro-bridge and the micro-cross ($\alpha=\beta=\gamma=90^\circ$) and no significant lattice rotation can be observed in these center regions, \textit{i.e.} the regions of interest for strain concentration. 
This confirms the ideal character of the strain concentration approach in the narrowest part of the micro-bridges and micro-crosses to get a homogenous, pure biaxial or uniaxial stress without shear component. 

\subsection{Experimental full strain tensor measurement}
\begin{figure}\label{Fig6}
\caption{
(a) Normalized intensity of several Bragg reflections in bulk Ge as a function of the angle of the diamond plate, showing localized dips due to the diffraction of the corresponding energy by the upstream diamond plate. (b) Energy of the Bragg reflections in the diamond plate (lines) and of selected Bragg reflection in bulk Ge (circles), as a function of the diamond angle (lines). Thicker lines indicate larger structure factors and thus stronger attenuation downstream.}
\includegraphics[width=1.\textwidth]{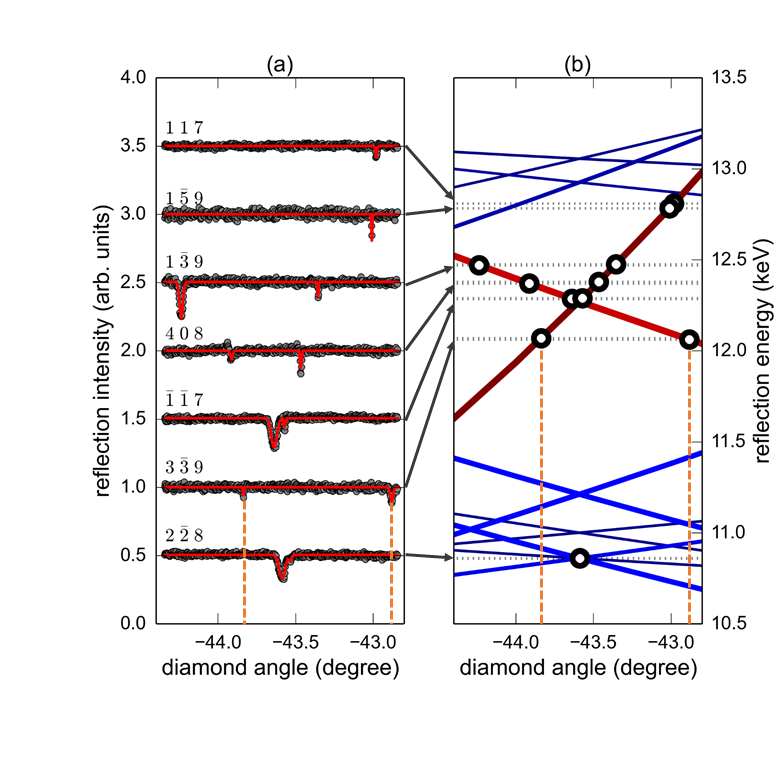}
\end{figure}

In order to access the full strain tensor from the deviatoric strain and the lattice tilts, we have performed rainbow-filtered Laue diffraction. 
The full strain tensor can be calculated from the combination of (\textit{i}) the deviatoric strain tensor measured using standard Laue micro-diffraction, and (\textit{ii}) at least one d-spacing obtained from the measurement of the energy (wavelength) of a Bragg diffraction peak. 
Indeed, all the lattice distances calculated from the deviatoric strain tensor are proportional to the actual strained lattice distances with the same constant scale factor, as described hereafter. 
The measurement of one actual lattice distance sets that scale factor. 
The advantages of measuring the d-spacings using the rainbow-filter technique is that the beam is not translated, as would be the case when using a double crystal or reflection monochromator.
Thus the optical axis remains the same during the Laue pattern collection and the deviatoric strain tensor, lattice tilts and d-spacings are measured at the same time.
The raw orientation and axis of rotation of the rainbow-filter, \textit{i.e.} the diamond plate, were measured using a secondary CCD detector. 
Then, calibration was performed using a $\left\langle111\right\rangle$ bulk Ge sample \cite{robach_tunable_2013}. 
The quality of the calibration was then verified by measuring several rainbow-filtered Bragg reflections in a bulk $\left\langle001\right\rangle$ germanium crystal, as shown in Fig. \ref{Fig6}(a). 
Several intensity drops can be observed on the selected Bragg reflections, corresponding to photons removed by upstream diffraction events in the diamond plate.
Since the orientation of the diamond plate is known for any rotation angle, all the energies corresponding to a diffraction condition in the diamond can be determined. 
These energies will be attenuated in the spectrum transmitted by the diamond and impinging on the sample. 
The lines in Figure \ref{Fig6}(b) display the corresponding energies as a function of the diamond plate angle (the thicker the line, the stronger the diffraction by the diamond plate and the stronger the downstream attenuation). 
Therefore, the rotation angle of the diamond plate at which a given Bragg reflection from the sample is attenuated (Fig. \ref{Fig6}(a)) gives the corresponding energy of that particular reflection (circles in Fig. \ref{Fig6}(b)). 
The corresponding lattice distance $d_{hkl}$ can then be calculated using Equation \ref{Eq1}.

\begin{equation}\label{Eq1}
d_{hkl} = \frac{E}{2hc\sin\theta}
\end{equation}
 
where $d_{hkl}$ is the distance between $hkl$ crystal planes, $E$ is the energy of the Bragg reflection (measured experimentally using the rainbow-filter technique), $h$ is the Planck constant, $c$ the celerity of light in vacuum and $\theta$ is the Bragg angle (the pixel position of the Laue peak on the detector gives the $2\theta$ value). 
Note that the intensity of a Bragg reflection from the sample may drop at different angles due to different diffractions from the diamond filter. 
For instance, both detected attenuations of the $3\bar{3}9$ reflection are indicated by the dashed vertical lines in Fig. \ref{Fig6}. 
The uncertainty evaluation in a microdiffraction experiment has been discussed in great details by Hofmann \textit{et al.} \cite{hofmann_analysis_2011}. In particular they have investigated the influence of the different detector parameters (detector position, tilts and pixel errors) on the experimental strain uncertainty, which was shown to be smaller than $5\times10^{-5}$ in optimum conditions. In the present case, the energy of the Bragg reflection (or correspondingly the $d_{hkl}$ value) is introduced as an additional parameter in the full strain tensor determination, as explained below. The energy determination is itself directly dependent on the measurement of the diamond filter angle at the attenuation. This measurement can be impaired by a low signal-to-noise ratio (for example for higher energy reflections that are less attenuated by the filter) and also by dynamical effects in the diamond plate that can result in an enhanced transmission. Additionally, the relationship between the attenuation angle and the energy relies on the quality of the calibration of the diamond filter. While it would certainly be possible to evaluate formally the resulting uncertainty, an experimental evaluation of the uncertainty can be obtained by simultaneous energy determination using different reflections from the filter. It is estimated to be about $0.05\%$ as explained below.

Table \ref{Tab1} summarizes the measured energies of the Bragg reflections from the sample ($hkl_{sample}$) from the attenuation due to the diffraction $hkl_{filter}$ of the diamond plate. 
Typical discrepancies between two energy determinations from two different diffractions from the filter are below 8~eV, \textit{i.e.} $\Delta E/E < 7\times10^{-4}$. 
We obtained $d_{hkl}$ values within about $0.05\%$ of the theoretical values expected for a perfect cubic crystal with lattice parameter $a_0 = 5.6575 \AA$ \cite{levinshtein_handbook_1996}, which is an estimate of our typical experimental lattice parameter uncertainty. Note that the lattice parameter uncertainty is set by the energy uncertainty and since it is larger than the typical deviatoric strain uncertainty, it dominates the total strain uncertainty.
More details on the rainbow-filter technique can be found in Ref. \cite{robach_tunable_2013}.

%

From the measurement of the Laue pattern, one obtains the relative lattice parameters $\frac{b}{a}$ and $\frac{c}{a}$, as well as the crystal angles $\alpha$, $\beta$ and $\gamma$.
In addition, the inverse distance between the $(hkl)$ planes in an arbitrary triclinic cystal can be calculated from the metric tensor and can be written as
%
%
%

\begin{equation}\label{Eq2.2}
\frac{1}{d_{hkl}^2} = \frac{1}{a^2} f_{hkl}\left(\frac{b}{a}, \frac{c}{a}, \alpha, \beta, \gamma\right),
\end{equation}

where
\begin{equation}\label{Eq2.3}
f_{hkl}\left(\frac{b}{a}, \frac{c}{a}, \alpha, \beta, \gamma\right) = \frac{1}{V'^2} \left(S'_{11}h^2+S'_{22}k^2+S'_{33}l^2+2S'_{23}kl+2S'_{13}hl+2S'_{12}hk \right),
\end{equation}

with
\begin{eqnarray*}
S'_{11} &=& S_{11}/a^4 = \left(\frac{b}{a}\right)^2 \left(\frac{c}{a}\right)^2 \sin^2 \alpha ,  \\
S'_{22} &=& S_{22}/a^4 = \left(\frac{c}{a}\right)^2         \sin^2 \beta  , \\
S'_{33} &=& S_{33}/a^4 = \left(\frac{b}{a}\right)^2         \sin^2 \gamma  , \\
S'_{23} &=& S_{23}/a^4 = \frac{b}{a} \frac{c}{a}     (\cos \beta \cos \gamma - \cos \alpha)  , \\
S'_{13} &=& S_{13}/a^4 = \left(\frac{b}{a}\right)^2 \frac{c}{a} (\cos \gamma \cos \alpha - \cos \beta)  , \\
S'_{12} &=& S_{12}/a^4 = \frac{b}{a} \left(\frac{c}{a}\right)^2 (\cos \alpha \cos \beta - \cos \gamma)  , \\
V'^2 &=& V^2/a^6 = \left(\frac{b}{a}\right)^2 \left(\frac{c}{a}\right)^2 \left( 1 - \cos^2 \alpha - \cos^2 \beta - \cos^2 \gamma + 2 \cos \alpha \cos \beta \cos \gamma \right) .
\end{eqnarray*}

Thus the lattice parameter $a$ can be directly obtained from the measurement of $\frac{b}{a}$, $\frac{c}{a}$, $\alpha$, $\beta$, $\gamma$ and $d_{hkl}$ using:

\begin{equation}\label{Eq2.4}
a = d_{hkl} \sqrt{f_{hkl}\left(\frac{b}{a}, \frac{c}{a}, \alpha, \beta, \gamma\right)} .
\end{equation}

The diagonal components of the full strain tensor are then simply given by:

\begin{equation}\label{Eq3}
\varepsilon_{aa} = \frac{a-a_0}{a_0},
\end{equation}

replacing $a$ by $b$ or $c$ accordingly. Note that the off-diagonal components of the full strain tensor are the same as those of the deviatoric strain tensor.

We now turn to the results in the micro-bridges shown in Fig.\ref{Fig2}. 
The rainbow-filter measurements of the strain tensor components in the center region of the bridges are reported in Table \ref{Tab2}. 
A minimum of two diffracted peaks have been measured in order to control the measurement reliability. 
The experimental uncertainties, estimated from two different Bragg reflections, are within $0.10\%$ and somewhat larger than for bulk Ge. 
A possible reason is the lower signal-to-noise ratio in the thin membranes. 
The small negative in-plane shear component $\varepsilon_{12}$ which increases with the longitudinal strain, is explained by a small in-plane misalignement between the axis of the micro-bridge and the crystallographic axis $a$. 
Accordingly, the out-of-plane shear components $\varepsilon_{23}$ and $\varepsilon_{13}$ are equal, as expected from the symmetry. 

While the rainbow-filter method provides an accurate measurement of the local full strain tensor, it is hardly suitable for strain mapping since the diamond plate must be scanned at each measurement position. 
In order to keep the mapping time within reasonable limits (\textit{i.e.} a few hours), we turned to a numerical determination of the full strain tensor from the Laue diffraction patterns.

\subsection{Full strain tensor calculation}
The full strain tensor was calculated for each measurement location from the deviatoric strain tensor using the following model. 
Since the micro-bridges and micro-crosses are suspended, their top and bottom surfaces can be considered free of normal stress. 
Using Voigt notation, the $\sigma_{33}$ component of the stress can consequently be set to zero in the second order generalized Hooke’s law:

\begin{equation}\label{eq:voigt}
\left( \begin{array}{c}
\sigma_{11} \\
\sigma_{22} \\
\sigma_{33}=0 \\
\sigma_{23} \\
\sigma_{13} \\
\sigma_{12} \end{array} \right) 
=
\left( \begin{array}{cccccc}
c_{11} & c_{12} & c_{12} &    0   &    0   &    0    \\
c_{12} & c_{11} & c_{12} &    0   &    0   &    0    \\
c_{12} & c_{12} & c_{11} &    0   &    0   &    0    \\
   0   &    0   &    0   & c_{44} &    0   &    0    \\
   0   &    0   &    0   &    0   & c_{44} &    0    \\
   0   &    0   &    0   &    0   &    0   & c_{44} \end{array} \right) 
\left( \begin{array}{c}
\varepsilon_{11} \\
\varepsilon_{22} \\
\varepsilon_{33} \\
\varepsilon_{23} \\
\varepsilon_{13} \\
\varepsilon_{12} \end{array} \right) .
\end{equation}

%

where the subscripts $1$, $2$, $3$ indicate the $\left[100\right]$, $\left[010\right]$, $\left[001\right]$ directions respectively, $\sigma$ is the stress tensor, $\varepsilon$ is the strain tensor and $c_{11}$, $c_{12}$, $c_{44}$ are the second order elastic constants in Ge, taken as 126 GPa, 44.0 GPa and 67.7 GPa, respectively \cite{levinshtein_handbook_1996}. 
In particular

\begin{equation}\label{Eq4}
\sigma_{33}=c_{12}\left(\varepsilon_{11}+\varepsilon_{22}\right)+c_{11}\varepsilon_{33}=0.
\end{equation}

The relation between the (full) strain tensor and the deviatoric strain tensor $\varepsilon'$ is

\begin{equation}\label{Eq5}
\varepsilon_{ij}=\varepsilon'_{ij}+\frac{\varepsilon_{h}}{3}\delta_{ij},
\end{equation}

where $\varepsilon_{h}$ is the hydrostatic strain. 
Thus, from Eq. \ref{Eq4} and \ref{Eq5} it follows that 

\begin{equation}\label{Eq6}
\varepsilon_{h}=3\frac{c_{12}\left(\varepsilon'_{11}+\varepsilon'_{22}\right)+c_{11}\varepsilon'_{33}}{2c_{12}+c_{11}}.
\end{equation}

\begin{figure}\label{Fig7}
\caption{
Longitudinal strain in a series of $\left\langle100\right\rangle$ micro-bridges measured using the rainbow-filter technique as a function of the longitudinal strain calculated with a $\sigma_{zz}=0$ hypothesis. The square symbols correspond to measurements in the micro-bridges described in the text and in Fig. \ref{Fig2}.
Various stretching arm lengths $L$ were probed, yielding different longitudinal strains.}
\includegraphics[width=1.\textwidth]{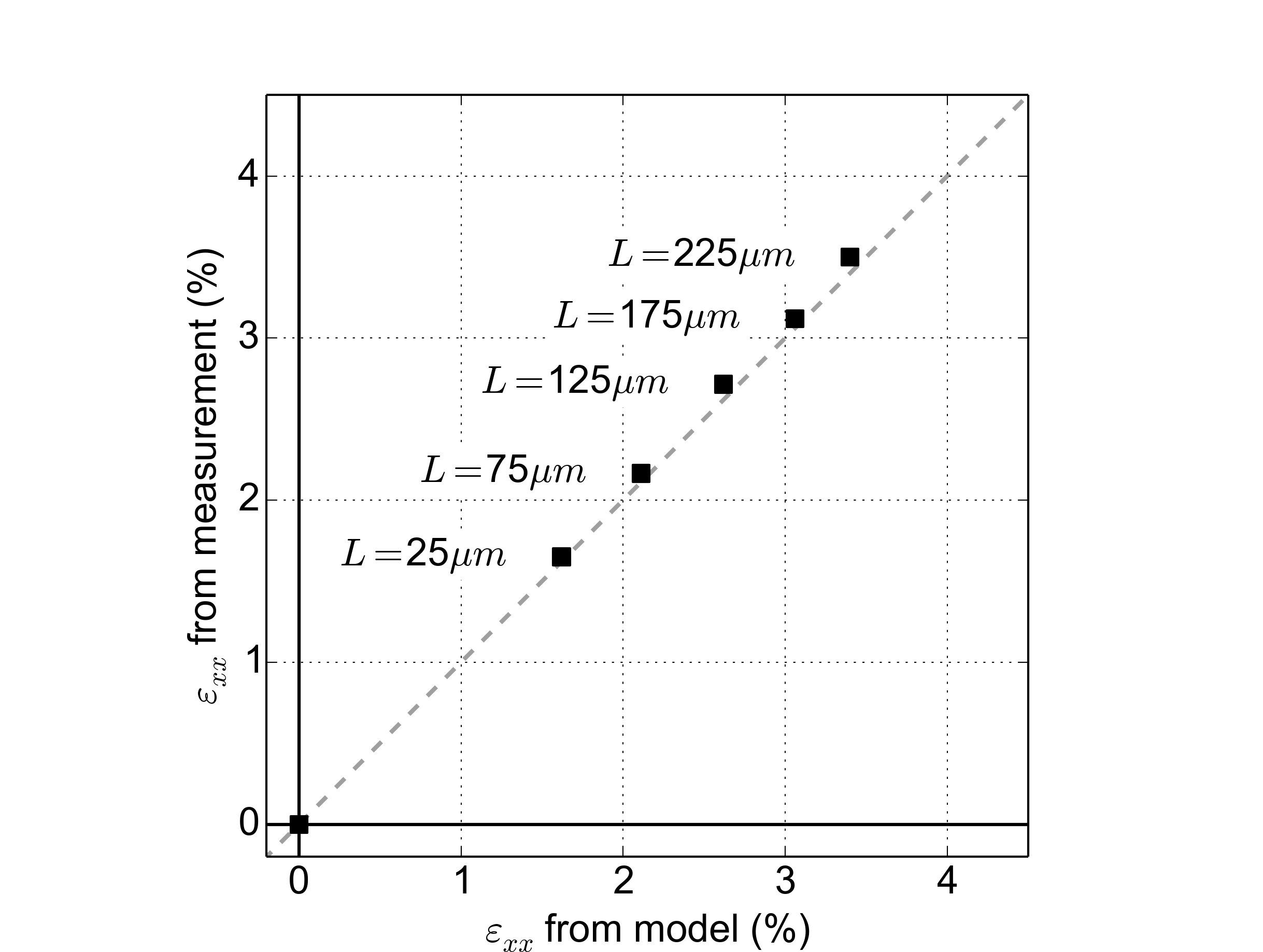}
\end{figure}

All the strain components can then be computed using Eq. \ref{Eq5}, Eq. \ref{Eq6}, and the measured values of the deviatoric strain tensor $\varepsilon'$. 
Furthermore, the remaining stress components can also be calculated from Eq. \ref{eq:voigt}. 
Our numerical approach was verified by confronting the measurements in the center region of micro-bridges using the rainbow-filter technique and the calculated strain tensor. 
The results for the longitudinal $\varepsilon_{11}=\varepsilon_{xx}$ component measured in different micro-bridges are shown in Fig. \ref{Fig7}. 
An excellent agreement is obtained, thus validating the numerical approach for faster full strain tensor mapping.
Furthermore, the agreement between the experimentally measured strain and the numerical calculation considering an elastic model confirms that there is no significant change in the value of the germanium elastic constants up to $5\%$ uniaxial strain.

\subsection{Full strain mapping}
\begin{figure}\label{Fig8}
\caption{
Components of the full strain tensor in a micro-bridge expressed in the $x = \left[100\right]$, $y = \left[010\right]$, $z = \left[001\right]$ basis, as obtained by FEM simulations (left) and rainbow-filtered Laue micro-diffraction measurements (right).}
\includegraphics[width=1.\textwidth]{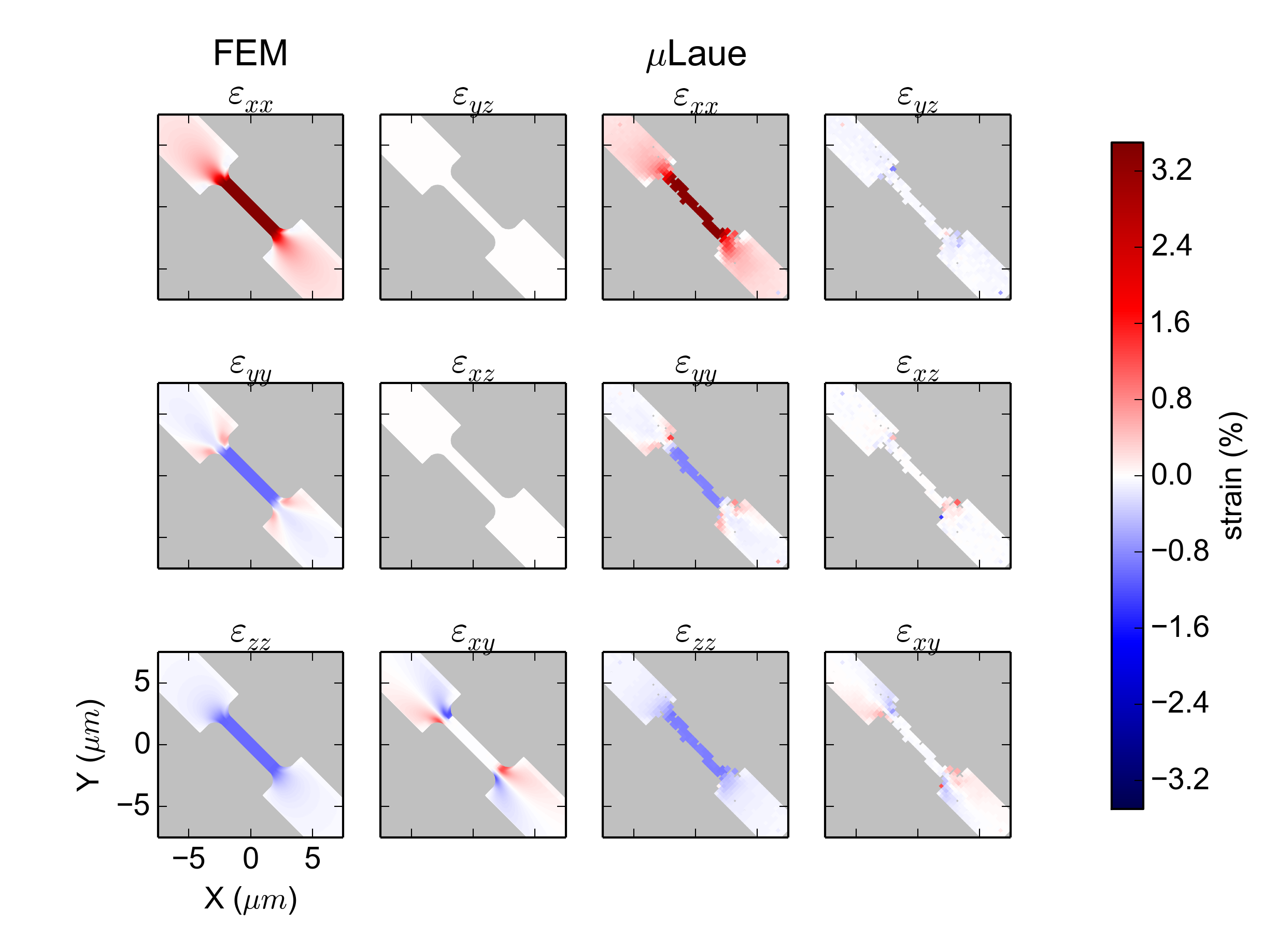}
\end{figure}

\begin{figure}\label{Fig9}
\caption{
Components of the full strain tensor in a micro-cross expressed in the $x = \left[110\right]$, $y = \left[\bar{1}10\right]$, $z = \left[001\right]$ basis, as obtained by FEM simulations (left) and rainbow-filtered Laue micro-diffraction measurements (right).}
\includegraphics[width=1.\textwidth]{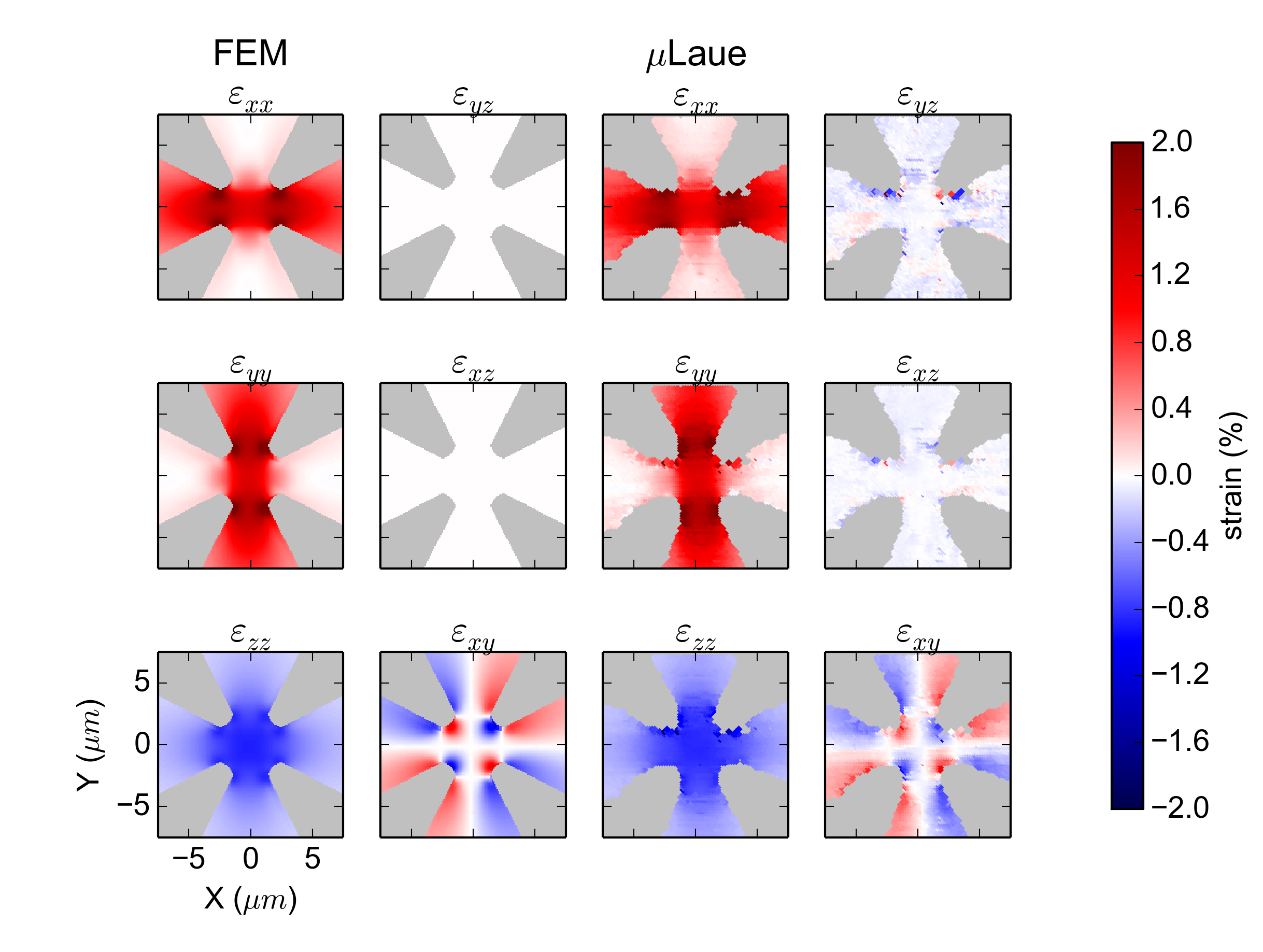}
\end{figure}

\begin{figure}\label{Fig10}
\caption{
Profile of the diagonal components of the full strain tensor across a micro-cross with $\left\langle110\right\rangle$ stretching arms, as measured experimentally with Laue micro-diffraction (symbols) and as calculated with FEM (solid lines).}
\includegraphics[width=1.\textwidth]{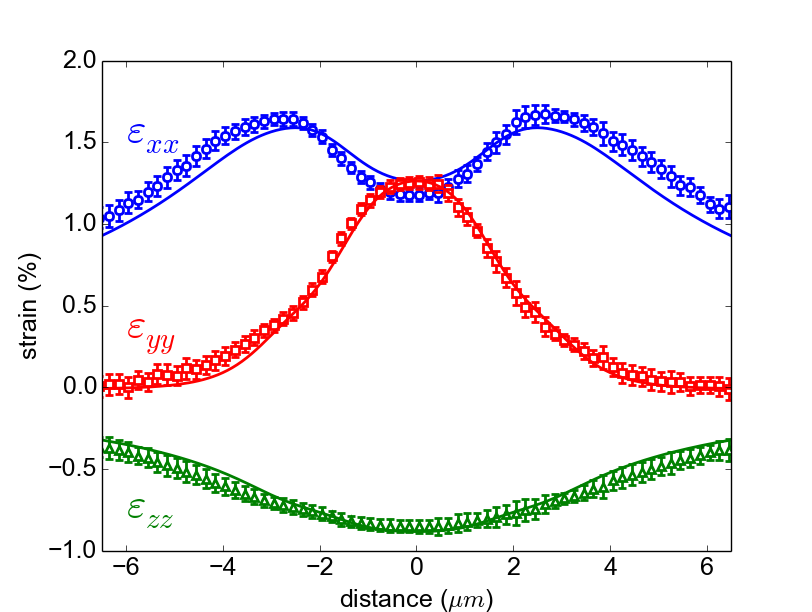}
\end{figure}

Using the numerical approach introduced before, we could calculate the full tensor from the measurements of the deviatoric strain tensor. 
The results are plotted in Fig. \ref{Fig8} and \ref{Fig9}, where the components of the full strain tensor are compared to FEM simulations results. 
An excellent agreement is obtained for all components in the micro-bridge, both in the center region and in the stretching pads (Fig. \ref{Fig8}). 
The agreement is also remarkable near the fillets where the strain varies rapidly and changes sign, even though the number of fitted peaks in these regions is lower and the average deviation of the fit is larger due to a less homogeneous strain along the beam path (not shown here). 
Figure \ref{Fig9} displays the results of FEM simulations and actual strain measurements in the micro-cross along $\left[110\right]$, $\left[\bar{1}10\right]$ and $\left[001\right]$, \textit{i.e.} along each stretching pad and perpendicularly to the membrane, respectively. 
The agreement is also very good: the center region is under uniform biaxial strain and the larger strain regions are located around the fillets in the pads. 
The simulated strain in the stretching pads away from the center region and the edges is uniaxial: the strain along the axis of the pad is positive; the out-of-plane component is negative and the in-plane strain perpendicular to the axis of the pad vanishes. 
The latter is expected since the Poisson ratio for the $\left[110\right]$ and $\left[\bar{1}10\right]$ directions is about ten times smaller than that for the $\left[100\right]$ and $\left[010\right]$ directions \cite{wortman_youngs_1965}. 
As mentionned earlier, the small out-of-plane shear strain components in the stretching arms of the micro-bridges and micro-crosses may be due small unetched layers or patches on the bottom surface.
Finally, figure \ref{Fig10} shows the profile of the three diagonal components across the micro-cross as measured using Laue micro-diffraction and as calculated with FEM, taking into account the beam shape. The errorbars of the experimental data have been estimated as follows: the experimental uncertainty is given by the sum of the deviatoric strain uncertainty $\Delta \varepsilon_d$ and the hydrostatic strain uncertainty $\Delta \varepsilon_h /3$. From equations \ref{Eq5} and \ref{Eq6}, it is clear that $\Delta \varepsilon_h = 3 \Delta \varepsilon_d$. As a result, the full strain uncertainty $\Delta \varepsilon$ can be estimated as $\Delta \varepsilon = \Delta \varepsilon_d + \Delta \varepsilon_h /3 = 2 f_{pix} \Delta_{pix}$, where $f_{pix}$ is the strain-to-pixel-error coefficient reported in Ref. \cite{hofmann_analysis_2011} and $\Delta_{pix}$ is the mean pixel deviation of the Laue fit of a given data point. Note that the experimental uncertainty stems mostly from the strain gradient, \textit{i.e.} it is smaller in the center region where the strain is homogenous and constant.
The agreement is remarkable for all three components simultaneously and we note that the small inconsistencies that exist between the FEM model and the actual measurement may be due to effects not captured by the model, such as the aforementioned small unetched patches of Si under the Ge micro-cross.

\section{Conclusion}
Using Laue diffraction measurements at the micrometer scale, we performed lattice distortions and tilts mapping in Ge micro-structures.
We observed homogeneous, pure tetragonal strain in the center regions and rapidly varying strain at the junction with the stretching pads, where lattice distortions were evidenced. 
In order to assess the full strain tensor in such devices, rainbow-filtered measurements have been performed and confirmed the absence of normal stress on the free surfaces, as well as the absence of any change in the value of the germanium elastic constants up to $5\%$ uniaxial strain.
As a result, full strain tensor maps could be directly calculated from standard Laue micro-diffraction measurements and comparison with FEM modeling showed a very good agreement. 
Therefore, our work demonstrates that the combination of the rainbow technique with Laue micro-diffraction leads to unambiguous strain and tilt mapping in crystalline strained micro-structures, such as GeOI micro-bridges or micro-crosses. 



\ack{Acknowledgements}

The authors would like to thank the ``Platforme de Technologie Amont'' in Grenoble for clean room facilities and the beamline BM32 at ESRF for synchrotron based measurements. 
This work was supported by the CEA DSM-DRT Phare projects ``Photonics'' and ``Operando'', the CEA-Enhanced Eurotalent project ``Straintronics'' as well as the Swiss National Science foundation.




\begin{table}\label{Tab1}
\caption{Selected Bragg reflections from the $\left\langle001\right\rangle$ bulk Ge sample $hkl_{sample}$ showing localized intensity drops due to the diffraction $hkl_{filter}$ from the upstream diamond plate filter. The energies can then be calculated and the corresponding lattice spacing $d_{hkl,exp}$ compared with the theoretical value $d_{hkl,theo}$ for a perfect cubic crystal with lattice parameter $5.6575\AA$.}
\begin{tabular}{ccccccc}      
\\ \hline \hline \\
 $hkl_{sample}$&$hkl_{filter}$&Filter angle&Energy&$d_{hkl,exp}$&$d_{hkl,theo}$ &$\frac{\Delta d_{hkl}}{d_{hkl}}$    \\
 & & ($^\circ$)& (keV)&($\AA$)&($\AA$)&($10^{-4}$)    \\
\\ \hline \\
$1 1 7$      	     & $1 \bar{1} 1$	            & $ -42.980$ & $	12.810$ & $	0.79229$ & $	0.79221$ & $	1.0 $ \\
\\ \hline \\
$1 \bar{5} 9$	     & $1 \bar{1} 1$	            & $ -43.009$ & $	12.783$ & $	0.54699$ & $	0.54693$ & $	1.1 $ \\
\\ \hline \\
$1 \bar{3} 9$	     & $\bar{1} \bar{1} \bar{1}$	& $ -44.238$ & $	12.471$ & $	0.59336$ & $	0.59307$ & $	4.8 $ \\
$1 \bar{3} 9$	     & $1 \bar{1} 1$	            & $ -43.354$ & $	12.475$ & $	0.59317$ & $	0.59307$ & $	1.6 $ \\
\\ \hline \\
$4 0 8$	             & $\bar{1} \bar{1} \bar{1}$	& $ -43.914$ & $	12.370$ & $	0.63295$ & $	0.63253$ & $	6.6 $ \\
$4 0 8$	             & $1 \bar{1} 1$	            & $ -43.466$ & $	12.378$ & $	0.63251$ & $	0.63253$ & $	-0.3 $ \\
\\ \hline \\
$\bar{1} \bar{1} 7$	 & $\bar{1} \bar{1} \bar{1}$	& $ -43.639$ & $	12.286$ & $	0.79262$ & $	0.79221$ & $	5.2 $ \\
$\bar{1} \bar{1} 7$	 & $1 \bar{1} 1$	            & $ -43.571$ & $	12.289$ & $	0.79245$ & $	0.79221$ & $	3.0 $ \\
\\ \hline \\
$3 \bar{3} 9$	     & $1 \bar{1} 1$	            & $ -43.836$ & $	12.069$ & $	0.56863$ & $	0.56860$ & $	0.5 $ \\
$3 \bar{3} 9$	     & $\bar{1} \bar{1} \bar{1}$    & $ -42.881$ & $	12.062$ & $	0.56894$ & $ 0.56860$ & $	6.0 $ \\
\\ \hline \\
$2 \bar{2} 8$	     & $0 0 4$	                    & $ -43.587$ & $	10.861$ & $	0.66667$ & $	0.66674$ & $	-1.0 $ \\
$2 \bar{2} 8$	     & $\bar{4} 0 0$	            & $ -43.587$ & $	10.860$ & $	0.66677$ & $	0.66674$ & $	0.4 $ \\
\\ \hline \hline
\end{tabular}
\end{table}

\begin{table}\label{Tab2}
\caption{Components of the full strain tensor measured using the rainbow-filter method.  Bragg$_{RF}$ indicates the Bragg reflections from the samples used for the rainbow-filter method, $N_{fit}$ is the total number of Bragg reflections fitted in the Laue pattern and Dev is the mean deviation of the fit of the Laue pattern.}
\begin{tabular}{cccccccccc}      
\\ \hline \hline \\
$L$&Bragg$_{RF}$&$\varepsilon_{11}$&$\varepsilon_{22}$&$\varepsilon_{33}$&$\varepsilon_{23}$&$\varepsilon_{13}$&$\varepsilon_{12}$&$N_{fit}$&Dev\\
($\mu m$)& &($\%$)&($\%$)&($\%$)&($\%$)&($\%$)&($\%$)&(peaks)&(pixels)\\
\\ \hline \\
Ge & \footnote{For the bulk Ge, the results are the average of the 12 measurements from Table \ref{Tab1}.}         &0.03& 0.02& 0.03&0.01&0.01&-0.01&116&0.187\\
\\ \hline \\
25 &$\bar{4}26$&1.60&-0.42&-0.48&0.14&0.14&-0.11& 71&0.456\\
   &$\bar{4}08$&1.70&-0.32&-0.38&    &    &     &   &     \\
\\ \hline \\
75 &$\bar{4}26$&2.14&-0.51&-0.58&0.12&0.11&-0.11& 78&0.448\\
   &$\bar{4}08$&2.19&-0.46&-0.53&    &    &     &   &     \\
\\ \hline \\
125&$\bar{4}26$&2.70&-0.62&-0.70&0.11&0.12&-0.12& 79&0.457\\
   &$\bar{4}08$&2.73&-0.58&-0.66&    &    &     &   &     \\
\\ \hline \\
175&$\bar{4}26$&3.16&-0.74&-0.83&0.10&0.10&-0.13& 83&0.359\\
   &$1\bar{3}7$&3.08&-0.81&-0.90&    &    &     &   &     \\
\\ \hline \\
225&$\bar{4}26$&3.53&-0.84&-0.91&0.11&0.12&-0.15& 88&0.366\\
   &$1\bar{3}7$&3.47&-0.89&-0.96&    &    &     &   &     \\
\\ \hline \hline
\end{tabular}
\end{table}


\begin{thebibliography}{58}
\baselineskip=9pt\parskip=0pt

\harvarditem[Aldaghri \emph{et~al.}]{Aldaghri, Ikonic \harvardand\
  Kelsall}{2012}{aldaghri_optimum_2012}
Aldaghri, O., Ikonic, Z. \harvardand\ Kelsall, R.~W.  \harvardyearleft
  2012\harvardyearright{}.
\newblock \emph{Journal of Applied Physics}, \volbf{111}(5), 053106.

\harvarditem[Boztug \emph{et~al.}]{Boztug, S{\'a}nchez-P{\'e}rez, Sudradjat,
  Jacobson, Paskiewicz, Lagally \harvardand\
  Paiella}{2013}{boztug_tensilely_2013}
Boztug, C., S{\'a}nchez-P{\'e}rez, J.~R., Sudradjat, F.~F., Jacobson, R.,
  Paskiewicz, D.~M., Lagally, M.~G. \harvardand\ Paiella, R.  \harvardyearleft
  2013\harvardyearright{}.
\newblock \emph{Small}, \volbf{9}(4), 622--630.

\harvarditem[Capellini \emph{et~al.}]{Capellini, Kozlowski, Yamamoto, Lisker,
  Wenger, Niu, Zaumseil, Tillack, Ghrib, de~Kersauson, El~Kurdi, Boucaud
  \harvardand\ Schroeder}{2013}{capellini_strain_2013}
Capellini, G., Kozlowski, G., Yamamoto, Y., Lisker, M., Wenger, C., Niu, G.,
  Zaumseil, P., Tillack, B., Ghrib, A., de~Kersauson, M., El~Kurdi, M.,
  Boucaud, P. \harvardand\ Schroeder, T.  \harvardyearleft
  2013\harvardyearright{}.
\newblock \emph{Journal of Applied Physics}, \volbf{113}(1).

\harvarditem[Cerdeira \emph{et~al.}]{Cerdeira, Buchenauer, Pollak \harvardand\
  Cardona}{1972}{cerdeira_stress_1972}
Cerdeira, F., Buchenauer, C.~J., Pollak, F.~H. \harvardand\ Cardona, M.
  \harvardyearleft 1972\harvardyearright{}.
\newblock \emph{Phys. Rev. B}, \volbf{5}, 580--593.

\harvarditem[Chahine \emph{et~al.}]{Chahine, Zoellner, Richard, Guha, Reich,
  Zaumseil, Capellini, Schroeder \harvardand\
  Sch{\'u}lli}{2015}{chahine_strain_2015}
Chahine, G.~A., Zoellner, M.~H., Richard, M.-I., Guha, S., Reich, C., Zaumseil,
  P., Capellini, G., Schroeder, T. \harvardand\ Sch{\'u}lli, T.~U.
  \harvardyearleft 2015\harvardyearright{}.
\newblock \emph{Applied Physics Letters}, \volbf{106}(7), 071902.

\harvarditem[Chen \emph{et~al.}]{Chen, Dejoie \harvardand\
  Wenk}{2012}{chen_unambigous_2012}
Chen, K., Dejoie, C. \harvardand\ Wenk, H.-R.  \harvardyearleft
  2012\harvardyearright{}.
\newblock \emph{Journal of Applied Crystallography}, \volbf{45}(5), 982--989.

\harvarditem{Chung \harvardand\ Ice}{1999}{chung_automated_1999}
Chung, J.-S. \harvardand\ Ice, G.~E.  \harvardyearleft 1999\harvardyearright{}.
\newblock \emph{Journal of Applied Physics}, \volbf{86}(9), 5249--5255.
\newblock 00227.

\harvarditem[Dejoie \emph{et~al.}]{Dejoie, Tamura, Kunz, Goudeau \harvardand\
  Sciau}{2015}{dejoie_complementary_2015}
Dejoie, C., Tamura, N., Kunz, M., Goudeau, P. \harvardand\ Sciau, P.
  \harvardyearleft 2015\harvardyearright{}.
\newblock \emph{Journal of Applied Crystallography}, \volbf{48}(5), 1522--1533.

\harvarditem[Dutt \emph{et~al.}]{Dutt, Sukhdeo, Nam, Vulovic, Yuan \harvardand\
  Saraswat}{2012}{dutt_roadmap_2012}
Dutt, B., Sukhdeo, D., Nam, D., Vulovic, B., Yuan, Z. \harvardand\ Saraswat, K.
   \harvardyearleft 2012\harvardyearright{}.
\newblock \emph{IEEE Photonics Journal}, \volbf{4}(5), 2002--2009.

\harvarditem[El~Kurdi \emph{et~al.}]{El~Kurdi, Bertin, Martincic, de~Kersauson,
  Fishman, Sauvage, Bosseboeuf \harvardand\
  Boucaud}{2010\emph{a}}{kurdi_control_2010}
El~Kurdi, M., Bertin, H., Martincic, E., de~Kersauson, M., Fishman, G.,
  Sauvage, S., Bosseboeuf, A. \harvardand\ Boucaud, P.  \harvardyearleft
  2010\emph{a}\harvardyearright{}.
\newblock \emph{Applied Physics Letters}, \volbf{96}(4).

\harvarditem[El~Kurdi \emph{et~al.}]{El~Kurdi, Fishman, Sauvage \harvardand\
  Boucaud}{2010\emph{b}}{kurdi_band_2010}
El~Kurdi, M., Fishman, G., Sauvage, S. \harvardand\ Boucaud, P.
  \harvardyearleft 2010\emph{b}\harvardyearright{}.
\newblock \emph{Journal of Applied Physics}, \volbf{107}(1), 013710.

\harvarditem[Etzelstorfer \emph{et~al.}]{Etzelstorfer, S{\"u}ess, Schiefler,
  Jacques, Carbone, Chrastina, Isella, Spolenak, Stangl, Sigg \harvardand\
  Diaz}{2014}{etzelstorfer_scanning_2014}
Etzelstorfer, T., S{\"u}ess, M.~J., Schiefler, G.~L., Jacques, V. L.~R.,
  Carbone, D., Chrastina, D., Isella, G., Spolenak, R., Stangl, J., Sigg, H.
  \harvardand\ Diaz, A.  \harvardyearleft 2014\harvardyearright{}.
\newblock \emph{Journal of Synchrotron Radiation}, \volbf{21}(1), 111--118.

\harvarditem[Gassenq \emph{et~al.}]{Gassenq, Guilloy, Osvaldo~Dias, Pauc,
  Rouchon, Hartmann, Widiez, Tardif, Rieutord, Escalante, Duchemin, Niquet,
  Geiger, Zabel, Sigg, Faist, Chelnokov, Reboud \harvardand\
  Calvo}{2015}{gassenq_bi_2015}
Gassenq, A., Guilloy, K., Osvaldo~Dias, G., Pauc, N., Rouchon, D., Hartmann,
  J.-M., Widiez, J., Tardif, S., Rieutord, F., Escalante, J., Duchemin, I.,
  Niquet, Y.-M., Geiger, R., Zabel, T., Sigg, H., Faist, J., Chelnokov, A.,
  Reboud, V. \harvardand\ Calvo, V.  \harvardyearleft 2015\harvardyearright{}.
\newblock \emph{Applied Physics Letters}, \volbf{107}(19).

\harvarditem[Ghrib \emph{et~al.}]{Ghrib, El~Kurdi, de~Kersauson, Prost,
  Sauvage, Checoury, Beaudoin, Sagnes \harvardand\
  Boucaud}{2013}{ghrib_tensile_2013}
Ghrib, A., El~Kurdi, M., de~Kersauson, M., Prost, M., Sauvage, S., Checoury,
  X., Beaudoin, G., Sagnes, I. \harvardand\ Boucaud, P.  \harvardyearleft
  2013\harvardyearright{}.
\newblock \emph{Applied Physics Letters}, \volbf{102}(22).

\harvarditem[Guilloy \emph{et~al.}]{Guilloy, Pauc, Gassenq, Gentile, Tardif,
  Rieutord \harvardand\ Calvo}{2015}{guilloy_tensile_2015}
Guilloy, K., Pauc, N., Gassenq, A., Gentile, P., Tardif, S., Rieutord, F.
  \harvardand\ Calvo, V.  \harvardyearleft 2015\harvardyearright{}.
\newblock \emph{Nano Letters}, \volbf{15}(4), 2429--2433.

\harvarditem[Hofmann \emph{et~al.}]{Hofmann, Abbey, Liu, Xu, Usher, Balaur
  \harvardand\ Liu}{2013}{hofmann_x-ray_2013}
Hofmann, F., Abbey, B., Liu, W., Xu, R., Usher, B.~F., Balaur, E. \harvardand\
  Liu, Y.  \harvardyearleft 2013\harvardyearright{}.
\newblock \emph{Nature Communications}, \volbf{4}, 2774.
\newblock 00005.

\harvarditem[Hofmann \emph{et~al.}]{Hofmann, Eve, Belnoue, Micha \harvardand\
  Korsunsky}{2011}{hofmann_analysis_2011}
Hofmann, F., Eve, S., Belnoue, J., Micha, J.-S. \harvardand\ Korsunsky, A.~M.
  \harvardyearleft 2011\harvardyearright{}.
\newblock \emph{Nuclear Instruments and Methods in Physics Research Section A:
  Accelerators, Spectrometers, Detectors and Associated Equipment},
  \volbf{660}(1), 130--137.
\newblock 00017.

\harvarditem[Huang \emph{et~al.}]{Huang, Barabash, Ice, Liu, Liu, Kai
  \harvardand\ Liaw}{2009}{huang_cyclic_2009}
Huang, E.-W., Barabash, R.~I., Ice, G.~E., Liu, W., Liu, Y.-L., Kai, J.-J.
  \harvardand\ Liaw, P.~K.  \harvardyearleft 2009\harvardyearright{}.
\newblock \emph{JOM}, \volbf{61}(12), 53--58.
\newblock 00011.

\harvarditem[Ibrahim \emph{et~al.}]{Ibrahim, Castelier, Palancher, Bornert,
  Car{\'{e}} \harvardand\ Micha}{2015}{ibrahim_laue_2015}
Ibrahim, M., Castelier, {\'{E}}., Palancher, H., Bornert, M., Car{\'{e}}, S.
  \harvardand\ Micha, J.-S.  \harvardyearleft 2015\harvardyearright{}.
\newblock \emph{Journal of Applied Crystallography}, \volbf{48}(4), 990--999.

\harvarditem{Ice \harvardand\ Larson}{2000}{ice_3d_2000}
Ice, G.~E. \harvardand\ Larson, B.~C.  \harvardyearleft
  2000\harvardyearright{}.
\newblock \emph{Advanced Engineering Materials}, \volbf{2}(10), 643--646.
\newblock 00089.

\harvarditem[Ice \emph{et~al.}]{Ice, Larson, Yang, Budai, Tischler, Pang,
  Barabash \harvardand\ Liu}{2005}{ice_polychromatic_2005}
Ice, G.~E., Larson, B.~C., Yang, W., Budai, J.~D., Tischler, J.~Z., Pang, J.
  W.~L., Barabash, R.~I. \harvardand\ Liu, W.  \harvardyearleft
  2005\harvardyearright{}.
\newblock \emph{Journal of Synchrotron Radiation}, \volbf{12}(2), 155--162.

\harvarditem{Ice \harvardand\ Pang}{2009}{ice_tutorial_2009}
Ice, G.~E. \harvardand\ Pang, J. W.~L.  \harvardyearleft
  2009\harvardyearright{}.
\newblock \emph{Materials Characterization}, \volbf{60}(11), 1191--1201.
\newblock 00032.

\harvarditem[Ike \emph{et~al.}]{Ike, Nakatsuka, Moriyama, Kurosawa, Taoka,
  Imai, Kimura, Tezuka \harvardand\ Zaima}{2015}{ike_characterization_2015}
Ike, S., Nakatsuka, O., Moriyama, Y., Kurosawa, M., Taoka, N., Imai, Y.,
  Kimura, S., Tezuka, T. \harvardand\ Zaima, S.  \harvardyearleft
  2015\harvardyearright{}.
\newblock \emph{Applied Physics Letters}, \volbf{106}(18).

\harvarditem[Keplinger \emph{et~al.}]{Keplinger, Grifone, Greil, Kriegner,
  Persson, Lugstein, {Tobias Sch{\"u}lli} \harvardand\
  Stangl}{2016}{keplinger_strain_2016}
Keplinger, M., Grifone, R., Greil, J., Kriegner, D., Persson, J., Lugstein, A.,
  {Tobias Sch{\"u}lli} \harvardand\ Stangl, J.  \harvardyearleft
  2016\harvardyearright{}.
\newblock \emph{Nanotechnology}, \volbf{27}(5), 055705.
\newblock 00000.

\harvarditem[Kirchlechner \emph{et~al.}]{Kirchlechner, Imrich, Liegl,
  Pörnbacher, Micha, Ulrich \harvardand\
  Motz}{2015}{kirchlechner_reversibility_2015}
Kirchlechner, C., Imrich, P.~J., Liegl, W., Pörnbacher, J., Micha, J.~S.,
  Ulrich, O. \harvardand\ Motz, C.  \harvardyearleft 2015\harvardyearright{}.
\newblock \emph{Acta Materialia}, \volbf{94}, 69--77.
\newblock 00003.

\harvarditem[Kirchlechner \emph{et~al.}]{Kirchlechner, Keckes, Micha
  \harvardand\ Dehm}{2011\emph{a}}{kirchlechner_in_2011}
Kirchlechner, C., Keckes, J., Micha, J.-S. \harvardand\ Dehm, G.
  \harvardyearleft 2011\emph{a}\harvardyearright{}.
\newblock \emph{Advanced Engineering Materials}, \volbf{13}(8), 837--844.

\harvarditem[Kirchlechner \emph{et~al.}]{Kirchlechner, Kiener, Motz, Labat,
  Vaxelaire, Perroud, Micha, Ulrich, Thomas, Dehm \harvardand\
  Keckes}{2011\emph{b}}{kirchlechner_dislocation_2011}
Kirchlechner, C., Kiener, D., Motz, C., Labat, S., Vaxelaire, N., Perroud, O.,
  Micha, J.-S., Ulrich, O., Thomas, O., Dehm, G. \harvardand\ Keckes, J.
  \harvardyearleft 2011\emph{b}\harvardyearright{}.
\newblock \emph{Philosophical Magazine}, \volbf{91}(7-9), 1256--1264.
\newblock 00024.

\harvarditem[Korsunsky \emph{et~al.}]{Korsunsky, Hofmann, Abbey, Song, Belnoue,
  Mocuta \harvardand\ Dolbnya}{2012}{korsunsky_analysis_2012}
Korsunsky, A.~M., Hofmann, F., Abbey, B., Song, X., Belnoue, J.~P., Mocuta, C.
  \harvardand\ Dolbnya, I.  \harvardyearleft 2012\harvardyearright{}.
\newblock \emph{International Journal of Fatigue}, \volbf{42}, 1--13.
\newblock 00005.

\harvarditem[Kunz \emph{et~al.}]{Kunz, Tamura, Chen, MacDowell, Celestre,
  Church, Fakra, Domning, Glossinger, Kirschman, Morrison, Plate, Smith,
  Warwick, Yashchuk, Padmore \harvardand\ Ustundag}{2009}{kunz_dedicated_2009}
Kunz, M., Tamura, N., Chen, K., MacDowell, A.~A., Celestre, R.~S., Church,
  M.~M., Fakra, S., Domning, E.~E., Glossinger, J.~M., Kirschman, J.~L.,
  Morrison, G.~Y., Plate, D.~W., Smith, B.~V., Warwick, T., Yashchuk, V.~V.,
  Padmore, H.~A. \harvardand\ Ustundag, E.  \harvardyearleft
  2009\harvardyearright{}.
\newblock \emph{Review of Scientific Instruments}, \volbf{80}(3), 035108.
\newblock 00090.

\harvarditem[Larson \emph{et~al.}]{Larson, Yang, Ice, Budai \harvardand\
  Tischler}{2002}{larson_three_2002}
Larson, B.~C., Yang, W., Ice, G.~E., Budai, J.~D. \harvardand\ Tischler, J.~Z.
  \harvardyearleft 2002\harvardyearright{}.
\newblock \emph{Nature}, \volbf{415}(6874), 887--890.
\newblock 00511.

\harvarditem[Leclere \emph{et~al.}]{Leclere, Cornelius, Ren, Davydok, Micha,
  Robach, Richter, Belliard \harvardand\ Thomas}{2015}{leclere_in_2015}
Leclere, C., Cornelius, T.~W., Ren, Z., Davydok, A., Micha, J.-S., Robach, O.,
  Richter, G., Belliard, L. \harvardand\ Thomas, O.  \harvardyearleft
  2015\harvardyearright{}.
\newblock \emph{Journal of Applied Crystallography}, \volbf{48}(1), 291--296.

\harvarditem[Levinshtein \emph{et~al.}]{Levinshtein, Rumyantsev \harvardand\
  Shur}{1996}{levinshtein_handbook_1996}
Levinshtein, M., Rumyantsev, S. \harvardand\ Shur, M. (eds.)  \harvardyearleft
  1996\harvardyearright{}.
\newblock \emph{Handbook {Series} on {Semiconductor} {Paramet}}.
\newblock World Scientific Pub Co Inc.

\harvarditem[Lunt \emph{et~al.}]{Lunt, Baimpas, Salvati, Dolbnya, Sui, Ying,
  Zhang, Kleppe, Dluhoš \harvardand\ Korsunsky}{2015}{lunt_state_2015}
Lunt, A.~J., Baimpas, N., Salvati, E., Dolbnya, I.~P., Sui, T., Ying, S.,
  Zhang, H., Kleppe, A.~K., Dluhoš, J. \harvardand\ Korsunsky, A.~M.
  \harvardyearleft 2015\harvardyearright{}.
\newblock \emph{The Journal of Strain Analysis for Engineering Design},
  \volbf{50}(7), 426--444.
\newblock 00005.

\harvarditem[Maa{\ss} \emph{et~al.}]{Maa{\ss}, Grolimund, Van~Petegem,
  Willimann, Jensen, Van~Swygenhoven, Lehnert, Gijs, Volkert, Lilleodden
  \harvardand\ Schwaiger}{2006}{maass_defect_2006}
Maa{\ss}, R., Grolimund, D., Van~Petegem, S., Willimann, M., Jensen, M.,
  Van~Swygenhoven, H., Lehnert, T., Gijs, M. a.~M., Volkert, C.~A., Lilleodden,
  E.~T. \harvardand\ Schwaiger, R.  \harvardyearleft 2006\harvardyearright{}.
\newblock \emph{Applied Physics Letters}, \volbf{89}(15), 151905.
\newblock 00065.

\harvarditem[MacDowell \emph{et~al.}]{MacDowell, Celestre, Tamura, Spolenak,
  Valek, Brown, Bravman, Padmore, Batterman \harvardand\
  Patel}{2001}{macdowell_submicron_2001}
MacDowell, A.~A., Celestre, R.~S., Tamura, N., Spolenak, R., Valek, B., Brown,
  W.~L., Bravman, J.~C., Padmore, H.~A., Batterman, B.~W. \harvardand\ Patel,
  J.~R.  \harvardyearleft 2001\harvardyearright{}.
\newblock \emph{Nuclear Instruments and Methods in Physics Research Section A:
  Accelerators, Spectrometers, Detectors and Associated Equipment},
  \volbf{467–468, Part 2}, 936--943.
\newblock 00151.

\harvarditem[Mermoux \emph{et~al.}]{Mermoux, Crisci, Baillet, Destefanis,
  Rouchon, Papon \harvardand\ Hartmann}{2010}{mermoux_strain_2010}
Mermoux, M., Crisci, A., Baillet, F., Destefanis, V., Rouchon, D., Papon, A.~M.
  \harvardand\ Hartmann, J.~M.  \harvardyearleft 2010\harvardyearright{}.
\newblock \emph{Journal of Applied Physics}, \volbf{107}(1).

\harvarditem{Micha \harvardand\ Robach}{2010}{lauetools}
Micha, J.-S. \harvardand\ Robach, O.,  \harvardyearleft
  2010\harvardyearright{}.
\newblock {LaueTools} laue x-ray microdiffraction analysis software.
\newblock
  \url{http://www.esrf.eu/home/UsersAndScience/Experiments/CRG/BM32/lauetools--laue-x-ray-microdiffraction-analysis-software.html}.
\newblock Accessed: 2016-02-13.

\harvarditem[Minamisawa \emph{et~al.}]{Minamisawa, S{\"u}ess, Spolenak, Faist,
  David, Gobrecht, Bourdelle \harvardand\ Sigg}{2012}{minamisawa_top_2012}
Minamisawa, R., S{\"u}ess, M., Spolenak, R., Faist, J., David, C., Gobrecht,
  J., Bourdelle, K. \harvardand\ Sigg, H.  \harvardyearleft
  2012\harvardyearright{}.
\newblock \emph{Nat Commun}, \volbf{3}, 1096.

\harvarditem[Nam \emph{et~al.}]{Nam, Sukhdeo, Kang, Petykiewicz, Lee, Jung,
  Vučković, Brongersma \harvardand\ Saraswat}{2013}{nam_strain_2013}
Nam, D., Sukhdeo, D.~S., Kang, J.-H., Petykiewicz, J., Lee, J.~H., Jung, W.~S.,
  Vučković, J., Brongersma, M.~L. \harvardand\ Saraswat, K.~C.
  \harvardyearleft 2013\harvardyearright{}.
\newblock \emph{Nano Letters}, \volbf{13}(7), 3118--3123.

\harvarditem[Reboud \emph{et~al.}]{Reboud, Gassenq, Guilloy, Osvaldo~Dias,
  Escalante, Tardif, Pauc, Hartmann, Widiez, Gomez, Bellet~Amalric, Fowler,
  Rouchon, Duchemin, Niquet, Rieutord, Faist, Geiger, Zabel, Sigg, Chelnokov
  \harvardand\ Calvo}{2016}{reboud_ultra-high_2016}
Reboud, V., Gassenq, A., Guilloy, K., Osvaldo~Dias, G., Escalante, J., Tardif,
  S., Pauc, N., Hartmann, J.-M., Widiez, J., Gomez, E., Bellet~Amalric, E.,
  Fowler, D., Rouchon, D., Duchemin, Y., Niquet, Y.-M., Rieutord, F., Faist,
  J., Geiger, R., Zabel, T., Sigg, H., Chelnokov, A. \harvardand\ Calvo, V.
  \harvardyearleft 2016\harvardyearright{}.
\newblock \emph{Proc. SPIE}, \volbf{9752}, 975214.

\harvarditem[Reboud \emph{et~al.}]{Reboud, Widiez, Hartmann, Osvaldo~Dias,
  Fowler, Chelnokov, Gassenq, Guilloy, Pauc, Calvo, Geiger, Zabel, Faist
  \harvardand\ Sigg}{2015}{reboud_structural_2015}
Reboud, V., Widiez, J., Hartmann, J.~M., Osvaldo~Dias, G., Fowler, D.,
  Chelnokov, A., Gassenq, A., Guilloy, K., Pauc, N., Calvo, V., Geiger, R.,
  Zabel, T., Faist, J. \harvardand\ Sigg, H.  \harvardyearleft
  2015\harvardyearright{}.
\newblock \emph{Proc. SPIE}, \volbf{9367}, 936714--936714--6.

\harvarditem[Richard \emph{et~al.}]{Richard, Palancher, Castelier, Micha,
  Gamaleri, Carlot, Rouquette, Goudeau, Martin, Rieutord, Piron \harvardand\
  Garcia}{2012}{richard_strains_2012}
Richard, A., Palancher, H., Castelier, {\'{E}}., Micha, J.-S., Gamaleri, M.,
  Carlot, G., Rouquette, H., Goudeau, P., Martin, G., Rieutord, F., Piron,
  J.~P. \harvardand\ Garcia, P.  \harvardyearleft 2012\harvardyearright{}.
\newblock \emph{Journal of Applied Crystallography}, \volbf{45}(4), 826--833.

\harvarditem[Robach \emph{et~al.}]{Robach, Micha, Ulrich, Geaymond, Sicardy,
  H{\"a}rtwig \harvardand\ Rieutord}{2013}{robach_tunable_2013}
Robach, O., Micha, J.-S., Ulrich, O., Geaymond, O., Sicardy, O., H{\"a}rtwig,
  J. \harvardand\ Rieutord, F.  \harvardyearleft 2013\harvardyearright{}.
\newblock \emph{Acta Crystallographica Section A}, \volbf{69}(2), 164--170.

\harvarditem[Robach \emph{et~al.}]{Robach, Micha, Ulrich \harvardand\
  Gergaud}{2011}{robach_full_2011}
Robach, O., Micha, J.-S., Ulrich, O. \harvardand\ Gergaud, P.  \harvardyearleft
  2011\harvardyearright{}.
\newblock \emph{Journal of Applied Crystallography}, \volbf{44}(4), 688--696.

\harvarditem[Rogan \emph{et~al.}]{Rogan, Tamura, Swift \harvardand\
  {\"U}st{\"u}ndag}{2003}{rogan_direct_2003}
Rogan, R.~C., Tamura, N., Swift, G.~A. \harvardand\ {\"U}st{\"u}ndag, E.
  \harvardyearleft 2003\harvardyearright{}.
\newblock \emph{Nature Materials}, \volbf{2}(6), 379--381.
\newblock 00065.

\harvarditem[S\'anchez-P\'erez \emph{et~al.}]{S\'anchez-P\'erez, Boztug, Chen,
  Sudradjat, Paskiewicz, Jacobson, Lagally \harvardand\
  Paiella}{2011}{sanchez-perez_direct-bandgap_2011}
S\'anchez-P\'erez, J.~R., Boztug, C., Chen, F., Sudradjat, F.~F., Paskiewicz,
  D.~M., Jacobson, R.~B., Lagally, M.~G. \harvardand\ Paiella, R.
  \harvardyearleft 2011\harvardyearright{}.
\newblock \emph{Proceedings of the National Academy of Sciences},
  \volbf{108}(47), 18893--18898.
\newblock 00096.

\harvarditem[S{\"u}ess \emph{et~al.}]{S{\"u}ess, Geiger, Minamisawa, Schiefler,
  Frigerio, Chrastina, Isella, Spolenak, Faist \harvardand\
  Sigg}{2013}{suess_analysis_2013}
S{\"u}ess, M.~J., Geiger, R., Minamisawa, R.~A., Schiefler, G., Frigerio, J.,
  Chrastina, D., Isella, G., Spolenak, R., Faist, J. \harvardand\ Sigg, H.
  \harvardyearleft 2013\harvardyearright{}.
\newblock \emph{Nature Photonics}, \volbf{7}(6), 466--472.

\harvarditem[Sukhdeo \emph{et~al.}]{Sukhdeo, Nam, Kang, Brongersma \harvardand\
  Saraswat}{2014}{sukhdeo_direct_2014}
Sukhdeo, D.~S., Nam, D., Kang, J.-H., Brongersma, M.~L. \harvardand\ Saraswat,
  K.~C.  \harvardyearleft 2014\harvardyearright{}.
\newblock \emph{Photon. Res.} \volbf{2}(3), A8--A13.

\harvarditem[Tamura \emph{et~al.}]{Tamura, Celestre, MacDowell, Padmore,
  Spolenak, Valek, Meier~Chang, Manceau \harvardand\
  Patel}{2002\emph{a}}{tamura_submicron_2002}
Tamura, N., Celestre, R.~S., MacDowell, A.~A., Padmore, H.~A., Spolenak, R.,
  Valek, B.~C., Meier~Chang, N., Manceau, A. \harvardand\ Patel, J.~R.
  \harvardyearleft 2002\emph{a}\harvardyearright{}.
\newblock \emph{Review of Scientific Instruments}, \volbf{73}(3), 1369--1372.

\harvarditem[Tamura \emph{et~al.}]{Tamura, Chung, Ice, Larson, Budai, Tischler
  \harvardand\ Yoon}{1999}{tamura_strain_1999}
Tamura, N., Chung, J.-S., Ice, G.~E., Larson, B.~C., Budai, J.~D., Tischler,
  J.~Z. \harvardand\ Yoon, M.  \harvardyearleft 1999\harvardyearright{}.
\newblock \volbf{563}.
\newblock 00029.

\harvarditem[Tamura \emph{et~al.}]{Tamura, MacDowell, Celestre, Padmore, Valek,
  Bravman, Spolenak, Brown, Marieb, Fujimoto, Batterman \harvardand\
  Patel}{2002\emph{b}}{tamura_high_2002}
Tamura, N., MacDowell, A.~A., Celestre, R.~S., Padmore, H.~A., Valek, B.,
  Bravman, J.~C., Spolenak, R., Brown, W.~L., Marieb, T., Fujimoto, H.,
  Batterman, B.~W. \harvardand\ Patel, J.~R.  \harvardyearleft
  2002\emph{b}\harvardyearright{}.
\newblock \emph{Applied Physics Letters}, \volbf{80}(20), 3724--3726.

\harvarditem[Tamura \emph{et~al.}]{Tamura, MacDowell, Spolenak, Valek, Bravman,
  Brown, Celestre, Padmore, Batterman \harvardand\
  Patel}{2003}{tamura_scanning_2003}
Tamura, N., MacDowell, A.~A., Spolenak, R., Valek, B.~C., Bravman, J.~C.,
  Brown, W.~L., Celestre, R.~S., Padmore, H.~A., Batterman, B.~W. \harvardand\
  Patel, J.~R.  \harvardyearleft 2003\harvardyearright{}.
\newblock \emph{Journal of Synchrotron Radiation}, \volbf{10}(2), 137--143.

\harvarditem[Ulrich \emph{et~al.}]{Ulrich, Biquard, Bleuet, Geaymond, Gergaud,
  Micha, Robach \harvardand\ Rieutord}{2011}{ulrich_new_2011}
Ulrich, O., Biquard, X., Bleuet, P., Geaymond, O., Gergaud, P., Micha, J.~S.,
  Robach, O. \harvardand\ Rieutord, F.  \harvardyearleft
  2011\harvardyearright{}.
\newblock \emph{Review of Scientific Instruments}, \volbf{82}(3).

\harvarditem[Ure\~na \emph{et~al.}]{Ure\~na, Olsen \harvardand\
  Raskin}{2013}{urena_raman_2013}
Ure\~na, F., Olsen, S.~H. \harvardand\ Raskin, J.-P.  \harvardyearleft
  2013\harvardyearright{}.
\newblock \emph{Journal of Applied Physics}, \volbf{114}(14).

\harvarditem[Villanova \emph{et~al.}]{Villanova, Maurice, Micha, Bleuet,
  Sicardy \harvardand\ Fortunier}{2012}{villanova_multiscale_2012}
Villanova, J., Maurice, C., Micha, J.-S., Bleuet, P., Sicardy, O. \harvardand\
  Fortunier, R.  \harvardyearleft 2012\harvardyearright{}.
\newblock \emph{Journal of Applied Crystallography}, \volbf{45}(5), 926--935.

\harvarditem[Wilkinson \emph{et~al.}]{Wilkinson, Britton, Jiang \harvardand\
  Karamched}{2014}{wilkinson_review_2014}
Wilkinson, A.~J., Britton, T.~B., Jiang, J. \harvardand\ Karamched, P.~S.
  \harvardyearleft 2014\harvardyearright{}.
\newblock \emph{IOP Conference Series: Materials Science and Engineering},
  \volbf{55}(1), 012020.
\newblock 00009.

\harvarditem{Wolf}{1996}{wolf_micro_1996}
Wolf, I.~D.  \harvardyearleft 1996\harvardyearright{}.
\newblock \emph{Semiconductor Science and Technology}, \volbf{11}(2), 139.

\harvarditem{Wortman \harvardand\ Evans}{1965}{wortman_youngs_1965}
Wortman, J.~J. \harvardand\ Evans, R.~A.  \harvardyearleft
  1965\harvardyearright{}.
\newblock \emph{Journal of Applied Physics}, \volbf{36}(1), 153.

\end{thebibliography}


\begin{references}


\referencelist 

\end{references}
\end{document}